# A Theoretical Investigation of Surface Roughness Scattering in Silicon Nanowire Transistors

Jing Wang[*], Eric Polizzi[**], Avik Ghosh[*], Supriyo Datta[*] and Mark Lundstrom[*]

[*]School of Electrical and Computer Engineering, Purdue University, West Lafayette, Indiana 47907, USA

[**]Department of Computer Sciences, Purdue University, West Lafayette, Indiana 47907, USA

## ABSTRACT

In this letter, we report a three-dimensional (3D) quantum mechanical simulation to investigate the effects of surface roughness scattering (SRS) on the device characteristics of Si nanowire transistors (SNWTs). We treat the microscopic structure of the Si/SiO$_2$ interface roughness directly by using a 3D finite element technique. The results show that 1) SRS reduces the electron density of states in the channel, which increases the SNWT threshold voltage, and 2) the SRS in SNWTs becomes more effective when more propagating modes are occupied, which implies that SRS is more important in planar metal-oxide-semiconductor field-effect-transistors with many transverse modes occupied than in small-diameter SNWTs with few modes conducting.

The silicon nanowire transistor (SNWT) is attracting broad attention as a promising structure for future electronics.[1,2] Therefore, understanding carrier transport in Si nanowires becomes increasingly important. Careful studies are needed to experimentally characterize transport in SWNTs, but it is also clear that a theoretical understanding is similarly important. In this letter, we present a theoretical exploration of the $Si/SiO_2$ interface roughness scattering, or surface roughness scattering (SRS)[3-5], in SNWTs.

It is well-known that scattering due to $Si/SiO_2$ interface roughness is important in planar silicon metal-oxide-semiconductor field-effect transistors (MOSFETs), and it is expected to be even more important in ultra-thin body silicon-on-insulator (UTBSOI) MOSFETs.[3] For bulk MOSFETs, electrons are confined at the $Si/SiO_2$ interface by an electrostatic potential well. Under high gate bias, the potential well is thin, electrons are confined very near the interface, SRS increases, and the effective mobility decreases. For UTBSOI MOSFETs, the confining potential is determined by the film thickness, and SRS can be enhanced by the roughness at the two interfaces.[3] In a SNWT, the channel is surrounded by the $Si/SiO_2$ interfaces, so one might expect SRS to dominate transport. We will show, however, that SRS may be less important in SNWTs than in planar devices because of the one-dimensional (1D) nature of the SNWT channel.

Roughness at the $Si/SiO_2$ interface affects carrier transport in the following ways. First, roughness produces a fluctuating oxide thickness, causing variations in the electrostatic potential inside the Si body and thus in the electron subband profile. In addition, roughness changes the thickness of the $Si/SiO_2$ quantum well, inducing additional fluctuations in the subband energy and also variations in the wavefunction shape. The subband energy fluctuations, induced by both these effects, lead to fluctuating elements in the diagonal terms of the device Hamiltonian[6,7] and act as a scattering potential (here we call it 'Type I SRS'). At the same time, slice-to-slice variations in the wavefunction shapes along the channel produce deformation and coupling elements in both diagonal and off-diagonal terms of the device Hamiltonian,[6,7] and consequently lower the transmission. (This effect has been named "wavefunction deformation scattering,"[8-10] and we call it 'Type II SRS' here). In this letter, we treat both effects without invoking perturbation theory by directly simulating the physically rough SNWT.

In our work, we use a continuum, effective-mass description, and perform three-dimensional (3D), self-consistent quantum simulations[6,7,11] to investigate SRS in small-diameter (~3nm) SNWTs. The simulated structure is a gate-all-around SNWT with a rectangular cross section and a [100] oriented channel (see Fig. 1). Following previous work on SRS[3-5], we assume an abrupt, randomly varying interface between the Si and SiO$_2$, parametrized by a root mean square (*rms*) amplitude and an autocovariance function.[12,13] The statistical nature of the roughness will depend on the nanowire fabrication methods and may differ considerably from that arising during the high temperature oxidation of a planar Si surface. Nevertheless, since our objective is to discuss general insights into the physics of SRS in SNWTs, we will employ the roughness parameters for a planar (100) Si/SiO$_2$ interface obtained from Ref. 12. Our use of a continuum level description may be questioned, but we believe that it is a useful first step that gives insight into how the magnitude and spatial coherence of potential fluctuations influence carrier transport.

The microscopic structure of the Si/SiO$_2$ interface roughness is implemented into the 3D simulator in the following procedure. We first discretize the simulation domain with a 3D finite element mesh[6,7,11]; each element is a triangular prism with a *2Å* height and edge length, comparable to the size of roughness at the (100) Si/SiO$_2$ interface.[12] Next, we generate a two-dimensional (2D) random distribution across the *whole* Si/SiO$_2$ interface (unfolding the four interfacial planes into a sheet) according to an exponential autocovariance function,[12]

$$C(x) = \Delta_m^2 e^{-\sqrt{2}x/L_m}, \qquad (1)$$

where $L_m$ is the correlation length, $\Delta_m$ is the *rms* fluctuation of the roughness and $x$ is the distance between two sampling points at the interface. Based on the 2D random distribution, the types of the elements at the Si/SiO$_2$ interfaces may be changed from Si to SiO$_2$ or reversely, to mimic the rough interfaces (see Fig. 1 b).

After the roughness is implemented, electron transport through the rough SNWT is simulated by using the non-equilibrium Green's function approach[14]. With a coupled mode space (CMS) representation,[6,7,11] the wavefunction deformation due to the roughness is treated. (The simulation methodology has been discussed in detail in Refs. 6 and 7.) To emphasize the role of SRS on electron transport, we do not include any other scattering mechanisms, so coherent transport is assumed inside the device. (Oscillations in the current due to quantum interference might be expected, but the averaging over a thermal distribution of wavelengths that occurs is sufficient to suppress them.) The length of the channel (10 nm) is long enough to ensure that sufficient averaging takes place so that sample specific effects are not observed. The simulated results for the rough SNWT are then compared with those for a device with the same geometrical parameters (e.g., nominal oxide thickness and Si body thickness) but smooth Si/SiO$_2$ interfaces. By doing this, the effects of SRS on SNWT device characteristics can be clearly identified.

Figure 2 plots the electron subband profile (left column) at the ON-state ($V_{GS}=V_{DS}=$ 0.4V) in the simulated SNWT with rough and smooth Si/SiO$_2$ interfaces. The corresponding transmission coefficients (right column) for both the rough and smooth SNWTs are also shown. Note that the modes are coupled in the simulation; we show them separately for illustrative purposes only. It is clearly seen in the Energy *vs.* X plot that the presence of the roughness significantly affects electron subbands, leading to Type I SRS as mentioned earlier. At the same time, the shape of the confined wavefunction also alters from slice to slice in the rough SNWT, which causes Type II SRS. To examine the significance of Type II SRS, we plot an Energy *vs.* Transmission curve (dot-dashed) for the rough SNWT calculated by the uncoupled mode space (UMS) approach[6], in which only the variations in the electron subbands (Type I SRS) are included while the deformation and coupling terms (Type II SRS) are discarded. The fact that the UMS approach significantly overestimates the transmission for the rough device infers that Type II SRS is of crucial importance and must be considered in the simulation.

From the Energy *vs.* Transmission plot, we find that the difference between the transmission curve for the rough SNWT and that for the smooth device becomes more and more noticeable as energy increases. This is because as energy increases, more subbands (modes) become conductive and the coupling between different modes efficiently reduces the transmission in the rough SNWT. In other words, SRS becomes more significant as more modes conduct. As we will show later, this effect has an important impact on the role of SRS on SNWT device characteristics.

Figure 3 plots the $I_{DS}$ *vs.* $V_{GS}$ curves in a semi-logarithmic scale for both the rough and smooth SNWTs. The results show that there is a distinct threshold voltage ($V_T$) increase caused by SRS – if we define $V_T$ as $I_{DS}(V_{GS} = V_T, V_{DS} = 0.4\text{V}) = 2 \times 10^{-7} \cdot W_{Si}$ (Amp), where $W_{Si}$ is in nm, the $V_T$ increment is ~0.03V for the roughness parameters we used ($L_m$=0.7nm and *rms*=0.14nm) and varies little from sample to sample. The increase in $V_T$ due to SRS was unexpected and the reason for it is as follows. Due to SRS, injections at low energies are blocked in the rough SNWT, which reduces the density-of-states (DOS) near the band edge (see the Fig. 3 inset). The lowered DOS near the band edge reduces the charge density in the subthreshold regime, and consequently increases $V_T$ in the rough SNWT. This effect would be modest in a conventional MOSFET with an energy-independent DOS above the band edge, but it becomes pronounced in a 1D wire with a singularity in the DOS at the band edge.

Finally, we explore the effects of SRS on the SNWT drain current above threshold. To do this, we compute a current ratio $\beta = I_{DS}^{Rough} / I_{DS}^{Smooth}$ at the *same* gate overdrive, $V_{GS}$-$V_T$, for both rough and smooth SNWTs. By comparing currents (rough *vs.* smooth) at the same gate overdrive, the effect of the $V_T$ increasing induced by SRS is removed. This allows us to examine whether the roughness can cause a significant reduction of SNWT ON-current by back-scattering. Fig. 4 shows the $\beta$ vs. gate overdrive curves for the SNWTs with different wire widths and roughness parameters. Several interesting phenomena are observed. First, all the simulated structures display a decreasing $\beta$ with an increasing gate overdrive. This occurs because more modes become conductive under higher gate bias, which, as described earlier, enhances SRS in the SNWTs.

Second, based on the roughness parameters ($L_m$=0.7nm and $rms$=0.14nm) obtained from Ref. 12, the SNWT with $W_{Si}$=3nm (solid) obtains a $\beta \approx 0.9$ at a typical ON-state condition (gate overdrive = 0.3V for a 0.4V supplied voltage). To explore the effects of $L_m$, two additional values (1.4nm for circles and 3.0nm for triangles) were examined. The results show that $\beta$ is insensitive to $L_m$, as expected from the averaging over a distribution of wavelengths that occurs at room temperature and high drain bias ($V_{DS}$=0.4V). In contrast, doubling the $rms$ (diamonds) clearly degrades $\beta$ at the same gate overdrive, indicating the importance of maintaining relatively smooth Si/SiO$_2$ interfaces for the high performance application of SNWTs.

Third, increasing the wire width reduces the strength of quantum confinement and thus increases the number of conducting modes in the SNWT. Our results (solid vs. dashed) clearly show that with a larger number of conducting modes in the wider ($W_{Si}$=9nm) SNWT, SRS is much stronger than in the narrower ($W_{Si}$=3nm) device. This observation also implies that SRS is more serious in a planar MOSFET, which can be viewed as a SNWT with a very large wire width.

In summary, we theoretically investigated SRS in SNWTs by performing a full 3D, self-consistent, quantum mechanical simulation. The microscopic structure of the Si/SiO$_2$ interface roughness was implemented into the simulator using the 3D finite element method. We found that 1) SRS reduces the electron density of states in the channel, which increases the SNWT threshold voltage, and 2) SRS in SNWTs becomes more serious when more propagating modes conduct, implying that SRS is more important in planar MOSFETs with many transverse modes occupied.

This work provides important insights into the nature of SRS in SNWTs and suggests that SRS may not be as important in nanowires as it is in conventional, planar MOSFETs.

This work was supported by the Semiconductor Research Corporation (SRC), and the Microelectronics Advanced Research Corporation (MARCO) focus center on Materials, Structures and Devices (MSD). The authors thank Jin Zhang, Min Chin Chai, Joseph Taylor and Prof. M. Alam at Purdue University for their sincere help.

**FIGURE CAPTIONS:**

FIG. 1 The schematic diagram of the simulated gate-all-round SNWT ((b) shows one slice of the wire with a specific interface roughness pattern). The source/drain doping concentration is $2\times10^{20}\,cm^{-3}$ and the channel is undoped. There is no source/drain overlap with the channel and the gate length is $L$=10nm. For the device with smooth Si/SiO$_2$ interfaces, the Si body thickness is $T_{Si}$=3nm, the wire width is $W_{Si}$=3nm, and the oxide thickness is 1nm. $V_S$, $V_D$, $V_G$ are the applied voltage biases on the source, drain and gate, respectively.

FIG. 2 The electron subband profile (for the (010) valleys) and the corresponding transmission coefficients for the simulated SNWT ($T_{Si}$=$W_{Si}$=3nm) with smooth and rough Si/SiO$_2$ interfaces. The roughness parameters used are $L_m$=0.7nm and $rms$=0.14nm.[12] The device is at the ON-state ($V_{GS}$=$V_{DS}$=0.4V), so the source and drain Fermi levels are equal to 0eV and -0.4eV, respectively.

FIG. 3 $I_{DS}$ vs. $V_{GS}$ curves for the simulated SNWT ($T_{Si}$=$W_{Si}$=3nm) with smooth (solid) and rough (dashed with symbols) Si/SiO$_2$ interfaces. ($V_{DS}$=0.4V). Three samples (triangles, crosses and circles) of the rough SNWT are generated based on the same roughness parameters ($L_m$=0.7nm and $rms$=0.14nm) but different random number seeds. The inset illustrates the reduction of electron DOS at low injection energies caused by SRS.

FIG. 4 Current ratio $\beta$ vs. gate overdrive curves for the simulated SNWTs with different wire widths ($W_{Si}$) and roughness parameters ($L_m$ and $rms$). At all the cases, the Si body thickness is fixed to be $T_{Si}$=3nm and the drain bias is $V_{DS}$=0.4V.

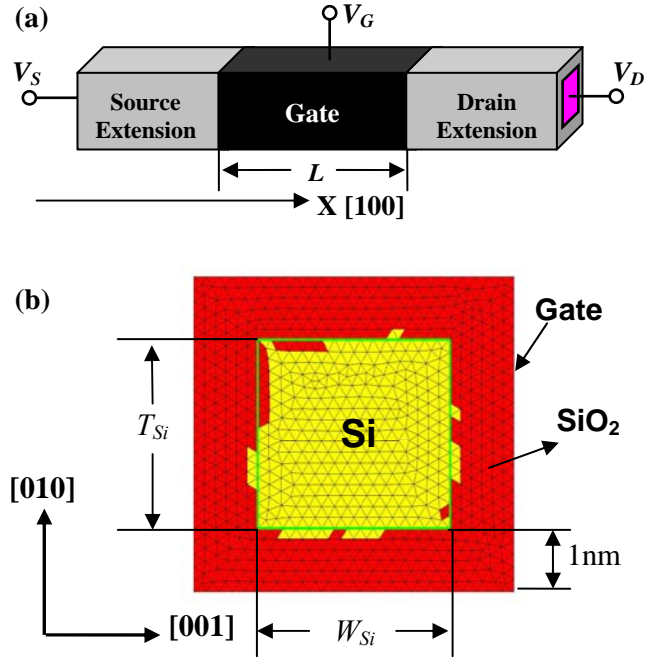

FIG. 1 The schematic diagram of the simulated gate-all-round SNWT ((b) shows one slice of the wire with a specific interface roughness pattern). The source/drain doping concentration is $2\times 10^{20}\,\mathrm{cm}^{-3}$ and the channel is undoped. There is no source/drain overlap with the channel and the gate length is $L=10$nm. For the device with smooth Si/SiO$_2$ interfaces, the Si body thickness is $T_{Si}=3$nm, the wire width is $W_{Si}=3$nm, and the oxide thickness is 1nm. $V_S$, $V_D$, $V_G$ are the applied voltage biases on the source, drain and gate, respectively.

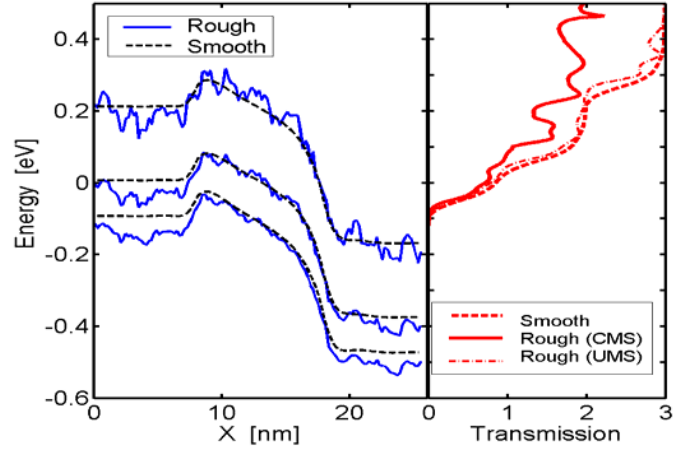

FIG. 2  The electron subband profile (for the (010) valleys) and the corresponding transmission coefficients for the simulated SNWT ($T_{Si}=W_{Si}=3$nm) with smooth and rough Si/SiO$_2$ interfaces. The roughness parameters used are $L_m=0.7$nm and $rms=0.14$nm.[12] The device is at the ON-state ($V_{GS}=V_{DS}=0.4$V), so the source and drain Fermi levels are equal to 0eV and -0.4eV, respectively.

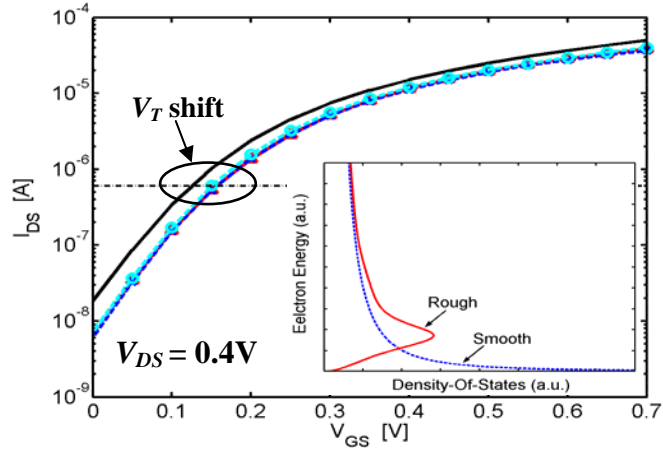

FIG. 3  $I_{DS}$ vs. $V_{GS}$ curves for the simulated SNWT ($T_{Si}=W_{Si}=3$nm) with smooth (solid) and rough (dashed with symbols) Si/SiO$_2$ interfaces. ($V_{DS}=0.4$V). Three samples (triangles, crosses and circles) of the rough SNWT are generated based on the same roughness parameters ($L_m=0.7$nm and $rms=0.14$nm) but different random number seeds. The inset illustrates the reduction of electron DOS at low injection energies caused by SRS.

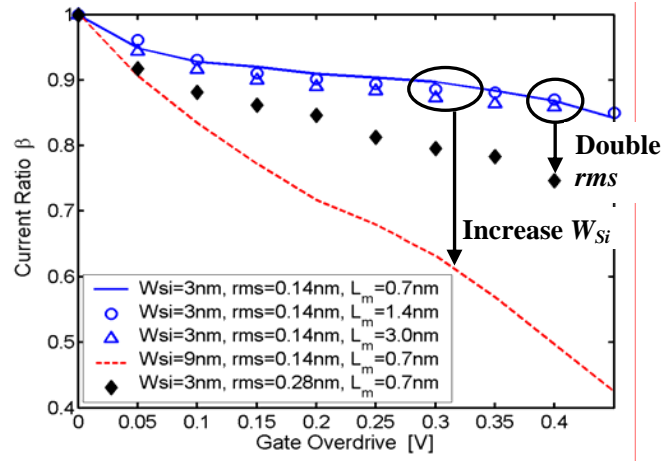

FIG. 4  Current ratio $\beta$ *vs.* gate overdrive curves for the simulated SNWTs with different wire widths ($W_{Si}$) and roughness parameters ($L_m$ and *rms*). At all the cases, the Si body thickness is fixed to be $T_{Si}$=3nm and the drain bias is $V_{DS}$=0.4V.